\documentstyle[12pt,prd,aps]{revtex}
\title{\bf{\Huge Maxwell-Boltzmann, Bose-Einstein and Fermi-Dirac statistical 
      entropies in a $D$-dimensional stationary axisymmetry space-time} }
\author{S. Q. Wu\thanks{E-mail: emu@iopp.ccnu.edu.cn}
   and X. Cai\thanks{E-mail: xcai@wuhan.cngb.com}\\
    \footnotesize \sl Institute of Particle Physics, Hua-Zhong
		      Normal University, Wuhan, 430079, China}

\begin{document}
\maketitle
\baselineskip 20pt
\begin{quote}

\ \ \ Statistical entropies of a general relativistic ideal gas obeying 
Maxwell-Boltzmann, Bose-Einstein and Fermi-Dirac statistics are calculated in a 
general axisymmetry space-time of arbitrary dimension. Analytical expressions 
for the thermodynamic potentials are presented, and their behaviors in the 
high or low temperature approximation are discussed. The entropy of a quantum 
field is shown to be proportional to the volume of optical space or that of 
the dragged optical space only in the high temperature approximation or in 
the zero mass case. In the case of a black hole, the entropy of a quantum 
field at the Hartle-Hawking temperature is proportional to the horizon "area" 
if and only if the horizon is located at the light velocity surface. 
 
\end{quote}

\vskip 8pt
\noindent
{\Large 1 Introduction}

Recently many efforts have been focused on understanding the statistical 
origin of black hole entropy$^1$ and its interpretation$^{2,3}$. One of them
is the so-called "brick-wall model" introduced by t' Hooft$^4$. By using 
this model, he first showed that the leading entropy of a quantum gas of 
scalar particles propagating outside the event horizon of the Schwarzschild 
black hole is proportional to its horizon area but diverges near the horizon. 
The divergences arise from the infinite one-particle number or state density 
of levels due to the presence of arbitrary high modes near the horizon. To
remove this divergence, t' Hooft introduced a brick wall cutoff which is
related to the horizon only.

Many works on the application of the brick-wall model to various kinds of
black hole have been done for scalar fields$^{4-11}$. In four dimensional 
Schwarzschild black hole, besides t' Hooft's pioneer work$^3$, the model has 
also been used to calculate the entropies for an idea gas obeying the usual 
three kinds of statistics$^5$. They showed that the area law for the entropy 
of a quantum field is due to the quantum statistics. Ghosh and Mitra$^6$ 
applied it to the Schwarzschild dilatonic black hole. In the case of a four 
dimensional rotating black hole, Lee and Kim$^{7,8}$ demonstrated that the 
leading term of the entropy of a neutral scalar field is proportional to the 
area of the event horizon and diverges as the system approaches to the event 
horizon. Similar conclusion holds true also for the leading entropy of a 
complex scalar field at the Hartle-Hawking temperature of a Kerr-Newman 
black hole$^9$. The reason of the divergences is attributed to the infinite 
number of state or the infinite volume of the phase space near the horizon. 
In other words, the origin of the divergence is that the density of states 
diverges at the horizon. In a $2+1$ dimensional black hole, the entropy of 
a quantum scalar field has also been studied in Ref. 10 and 11. 

It is well known that bosonic fields have a special class of superradiant mode 
solution$^{12}$. Because its complexity, the contribution to the entropy from 
superradiant modes hadn't been considered in Ref. 7-10. In the case of a Kerr
black hole, it is claimed in Ref. 13 that the negative contribution to the
entropy from superradiant modes is divergent in the leading order. However, Ho 
and Kang$^{14}$ pointed out that its entropy contribution is positive rather 
than negative and the previous error in Ref. 11 comes from the incorrect 
quantization of the superradiant modes. 
    
It is generally believed now that the entropy of a quantum field in a black 
hole is proportional to the area of a black hole horizon but diverges due to 
the presence of the event horizon. However, Alwis and Ohta$^{15}$ demonstrated 
that the free energy and the entropy for a scalar field or a Dirac field in 
the high temperature approximation is proportional to the volume of optical 
space in a static space-time background. Thus the horizon area of a black hole 
must have a certain relation to the volume of its corresponding optical space. 

Then the following questions arise: What relation between the entropy
calculated by the brick wall model method and that by the optical method?
Has any generality among the entropies of a quantum field in different
dimensional space-times or in different kinds of black holes with the same
dimensional number? Jing and Yan$^{16}$ had studied the entropy of a minimally
coupled quantum scalar field in the four dimensional general non-extremal
stationary axisymmetry black hole. However, for a general arbitrary dimensional 
space-time, there has no similar work presently. To fill up this gap, we 
investigate quantum statistics of a relativistic idea gas obeying three Kinds 
of the usual statistics: Maxwell-Boltzmann (M-B), Bose-Einstein (B-E) or 
Fermi-Dirac (F-D) in an arbitrary $D$-dimensional space-time.

In this paper, we first calculate the state density of a relativistical idea 
gas obeying three kinds of the usual statistics. We assume that this state 
density is effective for these three kinds of the usual statistics. Then we 
do thermodynamics in a curved space-time like the usual non-relativistical 
ones$^{17}$ in the flat space-time. The paper is organized as follows: In 
Sec. 2, we give a general description of a $D$-dimensional stationary 
axisymmetry space-time including static space-times and the flat space-time 
as well as black hole solutions. Then in Sec. 3, we derive the constraint 
on momentum space and evaluate the state density of phase space for a given 
energy. Sec. 4 and 5 is devoted to calculating thermodynamical potentials in 
two cases: the angular velocity of a quantum field $\Omega_0=0$ and $\Omega_0
\not=0$, respectively. Analytical expressions of the thermodynamical potentials 
are given, their asymptotical behaviors in the high or low temperature are 
discussed. General discussions about the divergence of the entropy of the 
quantum fields in a black hole background are presented in Sec. 6. Some four 
dimensional space-times are considered as examples in Sec. 7. Finally, we 
present our conclusions and problems not being considered here.
 
\vskip 8pt
\noindent
{\Large 2 Description of general space-time}

To begin with, let us firstly give some general description of a $D$-dimensional 
stationary axisymmetry space-time. As specially important examples, the flat 
space-time or a static space-time as well as black hole solutions are included 
under our present consideration.

In general, the line element and electro-magnetic potential of a $D$-dimensional 
stationary axisymmetry space-time can be expressed in the following form:
\begin{eqnarray}
ds^2&=&g_{tt}dt^2+2g_{t\varphi}dtd\varphi+g_{\varphi\varphi}d\varphi^2
+g_{ij}dx^idx^j, \\ 
A&=&A_tdt+A_{\varphi}d\varphi+A_idx^i, \hskip 0.2cm i,j=1,2,\cdots, D-2.
\end{eqnarray}

\noindent
where the metric elements $g_{tt}, g_{t\varphi}, g_{\varphi\varphi}$ and $g_{ij}$ 
are functions of coordinates $x=(x^1,x^2,\cdots,x^{D-2})$ only. The metric 
signature is taken as ($-,+,\cdots,+$) and its submatrix $g_{ij}$ is assumed 
to be diagonal. This space-time (1) has two Killing vectors $\partial_t$ and 
$\partial_{\varphi}$ relating to two conserved quantities, energy and azimuthal 
angular momentum. As examples, it contains a large class of black hole solutions 
and non-black-hole solutions such as the flat space-time. Static black holes 
are included as special case when $g_{t\varphi}=0$. The event horizon $f(x)=0$, 
if it exists, is a null hypersurface determined by
\begin{equation}
g^{tt}\partial_t^2f(x)+2g^{t\varphi}\partial_t\partial_{\varphi}f(x)+
g^{\varphi\varphi}\partial_{\varphi}^2f(x)+g^{ij}\partial_if(x)\partial_jf(x)
=g^{ij}\partial_if(x)\partial_jf(x)=0.
\end{equation}

\noindent
Here the contravariant metric elements being
\begin{equation} 
g^{tt}=g_{\varphi\varphi}/{\cal{D}}, g^{t\varphi}=-g_{t\varphi}/{\cal{D}},
g^{\varphi\varphi}=g_{tt}/{\cal{D}}, g^{ij}=1/g_{ij}, 
{\cal{D}}=g_{tt}g_{\varphi\varphi}-(g_{t\varphi})^2,
\end{equation}

\noindent
provided the metric determinant is nonsingular, namely, $g_D={\cal{D}}$
det$g_{ij}={\cal{D}}g_{D-2}\not=0$.

The condition that the nontrial null vector $N=(N_1,\cdots,N_{D-2})$ exists, 
where $N_i=\partial_if(x)$, is that the sub-determinant $g_{D-2}^{-1}$ of the 
contravariant metric tensor at the horizon ($r_h$) must be equal to zero, 
namely, $g_{D-2}^{-1}(r_h)=0$. As we assert that the metric determinant $g_D$ 
is non-singular, and so is the contravariant metric determinant $g_D^{-1}$. 
From the equality $g_{D-2}^{-1}={\cal{D}}g_D^{-1}$, one can know that the 
location of the horizon is given by the solutions of the following 
combination of equation and inequality:
\begin{equation}
g_{D-2}^{-1}(x)=0, \hskip 0.5cm{\rm namely}\hskip 0.5cm {\cal{D}}(x)=0,
\hskip 5pt {\rm and} \hskip 5pt g_D(x)\not=0.
\end{equation}

Eq.(2.3) in Ref. 18 is a part of this combination of equation and inequality
that determines the location of the horizon. For the metric component $g_{
\varphi\varphi}$ is nonzero at the event horizon ($r_h$), this expression is 
equivalent to Eq.(4) in Ref. 16, namely, $1/g^{tt}(r_h)=0$.  

In general, we can rewrite function ${\cal{D}}(x)$ around the point $r_h$ as
\begin{equation}
{\cal{D}}(x)=(r-r_h)^{\alpha}H(x)\approx(r-r_h)^{\alpha}H(r_h),
\end{equation}

\noindent
here $r$ is referred as to a radial coordinate in subspace, $H(x)$ is an 
analytical function at the point $r_h$, $\alpha$ is the order of zeros of 
function ${\cal{D}}(x)$, or the order of poles of the sub-determinant $g_{D-2}$.   
In other words, $\alpha$ is the number of duplicate roots of Eq.(5). For
the flat space-time, no solution to Eq.(5) exists, so the index $\alpha=0$; 
For a non-extremal black holes, Eq.(5) has a single root $r_h$, so $\alpha=1$; 
In the extremal case, it has a double root $r_h$, so we have $\alpha=2$, etc. 

\vskip 8pt
\noindent
{\Large 3 Deduction of momentum constraint and state density}

Before calculating the state density of single particle for a given energy 
from the volume of phase space, we firstly derive the constraint on momentum 
space. We proceed with the Lagrange-Hamiltonian method rather than from the 
semi-classical approximation of Klein-Gordon equation.

The Lagrangian of a relativistic charged particle in the above background 
space-time (1) is given by
\begin{equation}
2{\cal{L}}=g_{tt}\dot{t}^2+2g_{t\varphi}\dot{t}\dot{\varphi}
+g_{\varphi\varphi}\dot{\varphi}^2+g_{ij}\dot{x}^i\dot{x}^j+q(A_t\dot{t}
+A_{\varphi}\dot{\varphi}+A_i\dot{x}^i).
\end{equation}

Substituting the canonical conjugate momentum given by the following 
definition:
\begin{eqnarray}\nonumber
p_t&=&\frac{\partial{\cal{L}}}{\partial\dot{t}}=g_{tt}\dot{t}+g_{t\varphi}
\dot{\varphi}+qA_t=k_t+qA_t, \\  \nonumber
p_{\varphi}&=&\frac{\partial{\cal{L}}}{\partial\dot{\varphi}}=
g_{t\varphi}\dot{t}+g_{\varphi\varphi}\dot{\varphi}+qA_{\varphi}
=k_{\varphi}+qA_{\varphi}, \\  \nonumber
p_i&=&\frac{\partial{\cal{L}}}{\partial\dot{x}^i}=g_{ij}\dot{x}^j+qA_i
=k_i+qA_i,
\end{eqnarray}

\noindent
into the covariant Hamiltonian defined by ${\cal{H}}=p_t\dot{t}+p_{\varphi}
\dot{\varphi}+p_i\dot{x}^i-{\cal{L}}$, we get
\begin{equation}
2{\cal{H}}=g^{tt}(p_t-qA_t)^2+2g^{t\varphi}(p_t-qA_t)(p_{\varphi}-qA_{\varphi})
+g^{\varphi\varphi}(p_{\varphi}-qA_{\varphi})^2+g^{ij}(p_i-qA_i)(p_j-qA_j).
\end{equation}

Hamilton constraint $2{\cal{H}}=-\mu^2$ yields the constraint on momentum
space:
\begin{eqnarray}\nonumber
g^{tt}k_t^2+2g^{t\varphi}k_tk_{\varphi}+g^{\varphi\varphi}k_{\varphi}^2
+g^{ij}k_ik_j+\mu^2=g^{ij}(p_i-qA_i)(p_j-qA_j)+\mu^2  \\
+g^{tt}(\omega+qA_t)^2-2g^{t\varphi}(\omega+qA_t)(m-qA_{\varphi})
+g^{\varphi\varphi}(m-qA_{\varphi})^2=0,
\end{eqnarray}

\noindent
where we have put $p_t=-\omega, p_{\varphi}=m$, and $-k_t=\omega+qA_t$.

Let $p_t=\partial_t{\cal{S}}, p_{\varphi}=\partial_{\varphi}{\cal{S}}, p_i=
\partial_iS(x)$, where ${\cal{S}}=-\omega t+m\varphi+S(x)$, we can derive 
Hamilton-Jacobi (H-J) equation which is a rather good semi-classical (W-K-B) 
approximation to Klein-Gordon equation for a complex scalar field $\Psi=
e^{i{\cal{S}}}$ in this geometry.

General speaking, the H-J equation only has well-meaning for a scalar field. 
However, we assume that constraint Eq.(9) works for particles obeying Maxwell-
Boltzmann (M-B), Bose-Einstein (B-E) or Fermi-Dirac (F-D) statistics, and we 
will use it to calculate the density of single particle in the classical phase 
space. The reason is that the state density evaluated by the classical phase
space method is a rather good approximation to degeneracy of discrete levels 
in quantum case, while the latter is, in general, very difficult to be dealt 
with. 

Secondly, we evaluate the density of single particle state in the case that 
a quantum field has a vanishing angular velocity $\Omega_0=0$. (For case 
$\Omega_0\not=0$, see below). Momentum constraint of Eq.(9) can be recast 
into form
\begin{equation}
\frac{k_ik_j}{g_{ij}}+\frac{-g_{tt}}{-\cal{D}}[m-qA_{\varphi}
+\frac{g_{t\varphi}}{g_{tt}}(\omega+qA_t)]^2=
\frac{1}{-g_{tt}}(\omega+qA_t)^2-\mu^2,
\end{equation}

On the one hand, for a given energy $\omega$, the hypersurface represented 
by Eq.(10) is a ellipsoid in ($D-1$) dimensional momentum space supposed that 
it satisfies the following conditions: 
$$g_{ij}>0, \frac{-g_{tt}}{-\cal{D}}>0, \frac{(\omega+qA_t)^2}{-g_{tt}}>\mu^2.$$

To prevent from appearing an infinite and imaginary state density, these 
conditions must be satisfied. Thus we need to restrict the system in the 
region that $g_{tt}<0$. As the metric signature is taken as ($-,+,\cdots,+$), 
so we need $g_{ij}>0, -g_{tt}>0$, then the above conditions reduce to
\begin{equation}
-g_{tt}>0, \hskip 0.5cm -{\cal{D}}>0, \hskip 0.5cm 
\frac{(\omega+qA_t)^2}{-g_{tt}}>\mu^2.
\end{equation}

The first condition and second one place restriction on the coordinate space, 
the third one gives the lower bound on the range of energy. If these three 
conditions are satisfied, then the volume of phase space is finite. 
Otherwise, the hyper-surface is noncompact, and the state density $g(\omega)$ 
is divergent. The density of states for a given energy is given by taking 
differentiation of phase volume with respect to energy, $g(\omega)=d
\Gamma(\omega)/(d\omega)$, where $\Gamma(\omega)$ is the volume of $2(D-1)$ 
dimensional phase space for a given energy $\omega$: 
\begin{eqnarray}\nonumber
\Gamma(\omega)&=&\frac{1}{(2\pi)^{D-1}}\int d^{D-2}xd\varphi\int dmdk^{D-2} \\
&=&\frac{1}{(4\pi)^{\frac{D-1}{2}}\Gamma(\frac{D+1}{2})} 
\int d^{D-2}xd\varphi\sqrt{\frac{-g_D}{-g_{tt}}}
[\frac{(\omega+qA_t)^2}{-g_{tt}}-\mu^2]^{\frac{D-1}{2}}.
\end{eqnarray}

It must be emphasized that a factor $1/2$, though not important to the final
results, had been ignored in many literatures$^{4,6,8,10,11,13,14,16}$ in which 
the authors who used the radial wave number to calculate the free energy of a 
scalar field. The reason is that when one takes square roots of the radial 
wave number, he only takes a positive root and gets rid of a negative one. 
Physically, one discards the negative wave number; Mathematically, he gives 
up another leaf of paraboloid represented by the radial wave number. In the 
case of Minkowski space-time, the phase volume calculated by Eq.(12) is equal 
to the volume of the coordinate space times the volume of the momentum space 
being divided by a Planckian phase factor $(2\pi)^3$. Thus the total number of
single particle state computed from Eq.(12) is correct.

On the other hand, for a given energy $\omega$ and a fixed azimuthal angular 
momentum $m$, Eq.(9) represents a compact surface in ($D-2$) dimensional 
momentum space provided it satisfies the following angular momentum-energy 
constraint condition:
\begin{equation}
 g^{tt}(\omega+qA_t)^2-2g^{t\varphi}(\omega+qA_t)(m-qA_{\varphi})
+g^{\varphi\varphi}(m-qA_{\varphi})^2+\mu^2=-g^{ij}k_ik_j\leq 0,
\end{equation}

\noindent
due to $g_{ij}>0, k_i^2\geq 0$. Thus the volume of ($2D-3$) dimensional phase
space is easily computed,
\begin{eqnarray}\nonumber
\Gamma(\omega,m)=\frac{1}{(2\pi)^{D-2}}\int d^{D-2}xd\varphi\int dk^{D-2}=
\frac{1}{(4\pi)^{\frac{D-2}{2}}\Gamma(D/2)}\int d^{D-2}xd\varphi\sqrt{g_{D-2}}
\\  \times
[-g^{tt}(\omega+qA_t)^2+2g^{t\varphi}(\omega+qA_t)(m-qA_{\varphi}) 
-g^{\varphi\varphi}(m-qA_{\varphi})^2-\mu^2]^{\frac{D-2}{2}}.
\end{eqnarray}

The state density or the number of modes for a given $\omega$ and a fixed $m$ 
in ($2D-3$) dimensional space is given by $g(\omega,m)=d\Gamma(\omega,m)/(d
\omega)$. Using condition (13) and carrying out the integration or summation 
with respect to $m$, namely,
$$\Gamma(\omega)=\frac{1}{2\pi}\int dm\Gamma(\omega,m)
=\sum_m\Gamma(\omega,m), $$ 

\noindent
we can get the same result of the total number of state for a given energy 
$\omega$
$$g(\omega)=\frac{1}{2\pi}\int dm g(\omega,m)=\sum_mg(\omega,m).$$

\noindent
Here and after we take the quantum number $m$ as a continuous variable.

\vskip 8pt
\noindent
{\Large 4 Thermodynamical potential in the case ($\Omega_0=0$)}

The volume of phase space in Eq.(12) and its corresponding state density 
$g(\omega)$ are suitable to a quantum field that has a vanishing angular 
velocity but can have a potential $\Phi_0$. It is very convenient to use 
them in the case of static black holes and the flat space-time. We assume 
that a general relativistical bosonic, fermionic idea gas or non-interaction 
classical Boltzmann gas is in thermal equilibrium at temperature $1/\beta$ 
in the background space-time (1). The free energy for three kinds of the 
usual statistics is given by
\begin{eqnarray}
\beta F=\left\{\begin{array}{ll}
-\sum\limits_m\int d\omega g(\omega,m)e^{-\beta(\omega-q\Phi_0)}),&(M-B),\\
\sum\limits_m\int d\omega g(\omega,m)\ln(1-e^{-\beta(\omega-q\Phi_0)}),&(B-E),\\
-\sum\limits_m\int d\omega g(\omega,m)\ln(1+e^{-\beta(\omega-q\Phi_0)}),&(F-D).
\end{array}\right. 
\end{eqnarray}

After carrying the integration by parts on the r.h.s in Eq.(15) and making 
a substitution ${\cal{E}}=\omega+qA_t, {\cal{B}}=\Phi_0+A_t$, the above 
equation becomes
\begin{equation}
-F=\int d\omega\Gamma(\omega)
\left\{\begin{array}{l}
e^{-\beta(\omega-q\Phi_0)} \\
\frac{1}{e^{\beta(\omega-q\Phi_0)}-1}\\
\frac{1}{e^{\beta(\omega-q\Phi_0)}+1}
\end{array}\right. 
=\int d{\cal{E}}\Gamma({\cal{E}})
\left\{\begin{array}{ll}
e^{-\beta({\cal{E}}-q\cal{B})},&(M-B),\\
\frac{1}{e^{\beta({\cal{E}}-q\cal{B})}-1},&(B-E),\\
\frac{1}{e^{\beta({\cal{E}}-q\cal{B})}+1},&(F-D).
\end{array}\right. 
\end{equation}

The substitution ${\cal{E}}=-k_t=-p_t+qA_t=\omega+qA_t$ corresponds to a gauge 
transformation which doesn't alter the volume of phase space, thus the density
of state is invariant under such a gauge transformation. So we have $\Gamma(
\omega)=\Gamma({\cal{E}})$. One can alway choose such a gauge $\Phi_0$ that
makes ${\cal{B}}=\Phi_0+A_t=0$. Substituting the total number of state (namely,
phase volume) 
\begin{equation}
\Gamma({\cal{E}})=\frac{1}{(4\pi)^{\frac{D-1}{2}}\Gamma(\frac{D+1}{2})}
\int d^{D-2}xd\varphi\sqrt{\frac{-g_D}{-g_{tt}}}
[\frac{{\cal{E}}^2}{-g_{tt}}-\mu^2]^{\frac{D-1}{2}},
\end{equation}

\noindent
into the r.h.s of the second one in Eq.(16), we get the expression for the 
free energy
\begin{eqnarray}\nonumber
-F=\frac{1}{(4\pi)^{\frac{D-1}{2}}\Gamma(\frac{D+1}{2})} 
\int d^{D-2}xd\varphi\sqrt{\frac{-g_D}{-g_{tt}}} 
\int_{\mu\sqrt{-g_{tt}}}^{\infty}d{\cal{E}} \\ \times
[\frac{{\cal{E}}^2}{-g_{tt}}-\mu^2]^{\frac{D-1}{2}}
\left\{\begin{array}{ll}
e^{-\beta({\cal{E}}-q{\cal{B}})},&(M-B)\\
\frac{1}{e^{\beta({\cal{E}}-q{\cal{B}})}-1},&(B-E)\\
\frac{1}{e^{\beta({\cal{E}}-q{\cal{B}})}+1},&(F-D).
\end{array}\right. 
\end{eqnarray}

If $\mu\sqrt{-g_{tt}}>q{\cal{B}}$, then ${\cal{E}}>q{\cal{B}}$, in this case 
there exists no superradiant mode for a quantum bosonic field. However, if 
$\mu\sqrt{-g_{tt}}<q{\cal{B}}$, then there is an energy interval $\mu\sqrt{-
g_{tt}}\leq{\cal{E}}<q{\cal{B}}$ corresponding to the superradiant mode ($\omega
<q\Phi_0$) as well as another interval ${\cal{E}}>q{\cal{B}}$ corresponding 
to the non-superradiant mode ($\omega>q\Phi_0$). It is well known that there 
is no superradiant effect for a fermionic field. Because it is somewhat 
complicated in the superradiant case, we here shall not cope with it. Let us 
suppose ${\cal{E}}\geq\mu\sqrt{-g_{tt}}>q{\cal{B}}\geq 0$, then after some 
calculation, we get the final results for the free energy in this case. 
\begin{eqnarray}
-F=2(\frac{\mu}{2\pi})^{D/2}\int d^{D-2}xd\varphi
\frac{\sqrt{-g_D}}{(\beta\sqrt{-g_{tt}})^{D/2}}  
\left\{\begin{array}{ll}
e^{\beta q{\cal{B}}}K_{D/2}(\mu\beta\sqrt{-g_{tt}}),&(M-B),\\
\sum\limits_{n=1}^{\infty}\frac{e^{n\beta q{\cal{B}}}K_{D/2}(n\mu\beta
\sqrt{-g_{tt}})}{n^{D/2}},&(B-E),\\
\sum\limits_{n=1}^{\infty}\frac{e^{n\beta q{\cal{B}}}(-1)^{n+1}
K_{D/2}(n\mu\beta\sqrt{-g_{tt}})}{n^{D/2}},&(F-D).
\end{array}\right. 
\end{eqnarray}

\noindent
Here $K_{D/2}$ being the modified Bessel (or MacDonald) function of $D/2$-th 
order. 

By using the asymptotic expression of $D/2$-th order MacDonald function
$K_{D/2}(z)$ at small $z$:
$$K_{D/2}(z)\approx\frac{2^{\frac{D-2}{2}}\Gamma(D/2)}{z^{D/2}}, \hskip
0.5cm z\rightarrow 0$$

\noindent
one can obtain the asymptotic behavior of the free energy in the high 
temperature approximation ($\beta\rightarrow 0$) or in the zero mass case 
($\mu=0$) 
\begin{eqnarray}\nonumber
-F&\approx&\frac{\Gamma(D/2)}{\pi^{D/2}}\int d^{D-2}xd\varphi
\frac{\sqrt{-g_D}}{(\beta\sqrt{-g_{tt}})^D}
\left\{\begin{array}{ll}
e^{\beta q{\cal{B}}}\\
\sum\limits_{n=1}^{\infty}\frac{(e^{\beta q{\cal{B}}})^n}{n^D}\\
-\sum\limits_{n=1}^{\infty}\frac{(-e^{\beta q{\cal{B}}})^n}{n^D}
\end{array}\right. \\
&=&\frac{\Gamma(D/2)V_{D-1}}{\pi^{D/2}\beta^D}
\left\{\begin{array}{ll}
e^{\beta q{\cal{B}}},&(M-B)\\
\zeta_D(e^{\beta q{\cal{B}}}),&(B-E)\\
-\zeta_D(-e^{\beta q{\cal{B}}}),&(F-D)
\end{array}\right. 
\end{eqnarray}

\noindent
Here $\zeta_D(s)=\sum\limits_{n=1}^{\infty}\frac{s^n}{n^D}$ being Riemann
zeta function. $V_{D-1}=\int d^{D-2}xd\varphi\sqrt{\frac{-g_D}{(-g_{tt})^D}}$ 
is the volume of optical space$^{15}$ with metric determinant $\bar{g}_D=
g_D/(-g_{tt})^D$. The metric of optical space is
\begin{equation}
d\bar{s}^2=dt^2+2\frac{g_{t\varphi}}{g_{tt}}dtd\varphi
+\frac{g_{\varphi\varphi}}{g_{tt}}d\varphi^2+\frac{g_{ij}}{g_{tt}}dx^idx^j. 
\end{equation}

If we choose such a gauge potential $\Phi_0$ that makes ${\cal{B}}=0$, and 
introduce a convenient statistical factor ${\cal{N}}_D=1, \zeta_D(1), -\zeta_D(
-1)$, for M-B, B-E, F-D statistics, respectively, then the free energy will be
\begin{equation}
-F\approx{\cal{N}}_D\frac{\Gamma(D/2)V_{D-1}}{\pi^{D/2}\beta^D}.
\end{equation}

The free energy in Eq.(22) coincides with that calculated by the heat kernel 
expansion method in Ref. 15. In the case of a Dirac field, they only differs 
by a factor $2^{D/2}$ which is the number of a Dirac spinor components in a 
$D$-dimensional space and isn't considered by us here. In Ref. 15, the free 
energy is derived by functional integral method and made a heat kernel 
expansion in the high temperature approximation in the static space-time. 
However, the start-point of ours is a general stationary axisymmetry 
space-time.

The entropy is given by
\begin{equation}
S=\beta^2\frac{\partial F}{\partial\beta}
\approx{\cal{N}}_D\frac{D\Gamma(D/2)V_{D-1}}{\pi^{D/2}\beta^{D-1}}.
\end{equation}

The free energy in Eq.(22) and the entropy in Eq.(23) are proportional to
the volume of optical space and depend on the dimensional number $D$ of the
considered space-time. Their dependence on the space-time is only related to
the temporal component of the metric tensor. If the metric tensor $g_{tt}$
is nonzero everywhere in the space-time, there exists no divergence. However, 
if $g_{tt}$ vanishes at somewhere, then divergences appear there. Our results 
agree with that in Ref. 15 in the case of a four dimensional space-time. In 
the case of a black hole, only after introducing a brick wall cutoff and 
subtracting minor terms, can the entropy be proportional to the "area" of the 
event horizon at the Hartle-Hawking temperature $1/\beta_h=\kappa/(2\pi)$. 
(This will be illustrated below.)

\vskip 8pt
\noindent
{\Large 5 Thermodynamical potential in the case ($\Omega_0\not=0$)}

However, in the case that a quantum field has a nonzero angular velocity 
$\Omega_0$ in the stationary axisymmetry space-time, the matter is slightly 
different. Zhao and Gui$^{19}$ pointed out that "physical space" must be 
dragged by gravitational field with azimuthal angular velocity $\Omega_H$, 
and this is also noticed by Lee and Kim in Ref. 7-10 and other author$^{16}$. 
A classical relativistic idea gas or a quantum field in thermal equilibrium 
at temperature $1/\beta$ in this background must be dragged too. Therefore, 
it can be reasonable to assume that the quantum field or the classical 
particle is rotating with an azimuthal angular velocity $\Omega_0(x)$ and 
has a potential $\Phi_0(x)$. For such a modified angular momentum-energy 
equilibrium ensemble$^{14}$ of states of the field, the thermodynamic 
potential of the system for particles with charge $q$ and mass $\mu$ is 
given by
\begin{eqnarray}
\beta W=\left\{\begin{array}{ll}
-\sum\limits_m\int d\omega g(\omega,m)e^{-\beta(\omega
-m\Omega_0-q\Phi_0)}),&(M-B),\\
\sum\limits_m\int d\omega g(\omega,m)\ln(1-e^{-\beta(\omega
-m\Omega_0-q\Phi_0)}),&(B-E),\\
-\sum\limits_m\int d\omega g(\omega,m)\ln(1+e^{-\beta(\omega
-m\Omega_0-q\Phi_0)}),&(F-D).
\end{array}\right. 
\end{eqnarray}

Let us define $E-qB=\omega-m\Omega_0-q\Phi_0, B=\Phi_0+A_t+\Omega_0
A_{\varphi}$, then after carrying out the integration by parts in the 
r.h.s of Eq.(24), we obtain
\begin{eqnarray}
-W=\int dE\Gamma(E)
\left\{\begin{array}{ll}
e^{-\beta(E-qB)},&(M-B),\\
\frac{1}{e^{\beta(E-qB)}-1},&(B-E),\\
\frac{1}{e^{\beta(E-qB)}+1},&(F-D).
\end{array}\right. 
\end{eqnarray}

The total number of states $\Gamma(E)$ is obtained by substituting $\omega
+qA_t=E+\Omega_0(m-qA_{\varphi})$ into Eq.(9) and reducing this equation to 
Eq.(28) (see below). Now it is expressed as
\begin{eqnarray}
\Gamma(E)=\frac{1}{(4\pi)^{\frac{D-1}{2}}\Gamma(\frac{D+1}{2})}
\int d^{D-2}xd\varphi\sqrt{\frac{-g_D}{-\tilde{g}_{tt}}}
[\frac{E^2}{-\tilde{g}_{tt}}-\mu^2]^{\frac{D-1}{2}},
\end{eqnarray}

\noindent
here we have put $\tilde{g}_{tt}=g_{tt}+2g_{t\varphi}\Omega_0
+g_{\varphi\varphi}\Omega_0^2$.

The finiteness of the state density is guaranteed by the following conditions
\begin{equation}
-\tilde{g}_{tt}>0, \hskip 0.5cm -{\cal{D}}>0, \hskip 0.5cm \frac{E^2}{-
\tilde{g}_{tt}}>\mu^2
\end{equation}

\noindent
which comes from the compactness of hypersurface determined by
\begin{equation}
\frac{k_ik_j}{g_{ij}}+\frac{-\tilde{g}_{tt}}{-\cal{D}}[m-qA_{\varphi}
+\frac{g_{t\varphi}+g_{\varphi\varphi}\Omega_0}{\tilde{g}_{tt}}E]^2=
\frac{E^2}{-\tilde{g}_{tt}}-\mu^2.
\end{equation}

To preserve the state density real and finite, we must restrict the system in 
the region that satisfies $-\tilde{g}_{tt}>0$ because we want $-{\cal{D}}>0$ as
before. This imposes restrictions on the angular velocity in the region that 
$\Omega-\sqrt{-\cal{D}}/g_{\varphi\varphi}<\Omega_0<\Omega+\sqrt{-\cal{D}}/g_{
\varphi\varphi}$, where $\Omega=-g_{t\varphi}/g_{\varphi\varphi}$. Suppose that
$E\geq\mu\sqrt{-\tilde{g}_{tt}}>qB\geq 0$, namely, we only consider the case 
that the non-superradiant mode exists for a scalar field, as the calculation is 
somewhat complicated in the superradiant case. Substituting the total number of 
single particle state $\Gamma(E)$ into the thermodynamical potential, we have
\begin{eqnarray}\nonumber
-W=\frac{1}{(4\pi)^{\frac{D-1}{2}}\Gamma(\frac{D+1}{2})} 
\int d^{D-2}xd\varphi\sqrt{\frac{-g_D}{-\tilde{g}_{tt}}}
\int_{\mu\sqrt{-\tilde{g}_{tt}}}^{\infty}dE \\ \times
[\frac{E^2}{-\tilde{g}_{tt}}-\mu^2]^{\frac{D-1}{2}}
\left\{\begin{array}{ll}
e^{-\beta(E-qB)},&(M-B)\\
\frac{1}{e^{\beta(E-qB)}-1},&(B-E)\\
\frac{1}{e^{\beta(E-qB)}+1},&(F-D).
\end{array}\right. 
\end{eqnarray}

The thermodynamical potential for Bose-Einstein statistics coincides with 
Eq.(19) in Ref. 9 for a quantum scalar field in a four dimensional Kerr-Newman
black hole geometry. Under our assumption that $E\geq\mu\sqrt{-\tilde{g}_{tt}}
>qB\geq 0$, we can always select such a gauge potential $\Phi_0$ that makes $B=0$ 
without violating the above assumption. After carrying out the integration 
with respect to $E$, we arrive at results
\begin{eqnarray}
-W=2(\frac{\mu}{2\pi})^{D/2}\int d^{D-2}xd\varphi
\frac{\sqrt{-g_D}}{(\beta\sqrt{-\tilde{g}_{tt}})^{D/2}}
\left\{\begin{array}{ll}
K_{D/2}(\mu\beta\sqrt{-\tilde{g}_{tt}}),&(M-B),\\
\sum\limits_{n=1}^{\infty}\frac{K_{D/2}(n\mu\beta\sqrt{-
\tilde{g}_{tt}})}{n^{D/2}},&(B-E),\\
\sum\limits_{n=1}^{\infty}\frac{(-1)^{n+1}K_{D/2}(n\mu\beta\sqrt{-
\tilde{g}_{tt}})}{n^{D/2}},&(F-D).
\end{array}\right. 
\end{eqnarray}

In the high temperature approximation ($\beta\rightarrow 0$) or in the zero
mass case ($\mu=0$), the thermodynamical potential has asymptotic behavior: 
\begin{equation}
-W\approx\frac{\Gamma(D/2)}{\pi^{D/2}}\int d^{D-2}xd\varphi
\frac{\sqrt{-g_D}}{(\beta\sqrt{-\tilde{g}_{tt}})^D}{\cal{N}}_D
={\cal{N}}_D\frac{\Gamma(D/2)\tilde{V}_{D-1}}{\pi^{D/2}\beta^D}.
\end{equation}

\noindent
Here $\tilde{V}_{D-1}=\int d^{D-2}xd\varphi\sqrt{\frac{-g_D}{(-\tilde{g}_{tt}
)^D}}$ is the volume of the dragged optical space with determinant $\tilde{g}_D
=g_D/(-\tilde{g}_{tt})^D$. The metric of the dragged optical space is
\begin{equation}
d\tilde{s}^2=dt^2+2\frac{g_{t\varphi}+g_{\varphi\varphi}\Omega_0}{
\tilde{g}_{tt}}dtd\varphi+\frac{g_{\varphi\varphi}}{\tilde{g}_{tt}}
d\varphi^2+\frac{g_{ij}}{\tilde{g}_{tt}}dx^idx^j. 
\end{equation}

The entropy in the high temperature approximation is given by
\begin{equation}
S=\beta^2\frac{\partial W}{\partial\beta}
\approx{\cal{N}}_D\frac{D\Gamma(D/2)\tilde{V}_{D-1}}{\pi^{D/2}\beta^{D-1}}.
\end{equation}

The thermodynamical potential and its corresponding entropy are proportional
to the volume of the dragged optical space in the high temperature approximation. 
Apparently, they depend on the temperature and the dimensional number of the 
considered space-time as well as the temporal component of the dragged metric.
Their divergences count on the property of the dragged metric tensor $\tilde{g
}_{tt}$. No divergence will appear when $\tilde{g}_{tt}$ is nonzero everywhere. 
When the angular velocity $\Omega_0$ vanishes, they will degenerate to the case 
considered in the last section.

Using the asymptotic expression of $D/2$-th order MacDonald function $K_{D/2}
(z)$ at large $z$ and taking only its zero order approximation $$K_{D/2}(z)
\approx\sqrt{\frac{\pi}{2z}}e^{-z}, \hskip 0.5cm z\rightarrow \infty$$

\noindent
we can obtain the asymptotic behavior of the thermodynamical potential in
the low temperature approximation ($\beta\rightarrow\infty$):
\begin{eqnarray}\nonumber 
-W&\approx&(\frac{\mu}{2\pi})^{\frac{D-1}{2}}\int d^{D-2}xd\varphi
\frac{\sqrt{-g_D}}{(\beta\sqrt{-\tilde{g}_{tt}})^{\frac{D+1}{2}}}
\left\{\begin{array}{ll}
e^{-\beta\mu\sqrt{-\tilde{g}_{tt}}}\\
\sum\limits_{n=1}^{\infty}\frac{e^{-n\beta\mu\sqrt{-
\tilde{g}_{tt}}}}{n^{\frac{D+1}{2}}}\\
\sum\limits_{n=1}^{\infty}\frac{(-1)^{n+1}e^{-n\beta\mu\sqrt{-
\tilde{g}_{tt}}}}{n^{\frac{D+1}{2}}}
\end{array}\right. \\ \nonumber
&=&(\frac{\mu}{2\pi})^{\frac{D-1}{2}}\int d^{D-2}xd\varphi
\frac{\sqrt{-g_D}}{(\beta\sqrt{-\tilde{g}_{tt}})^{\frac{D+1}{2}}}
\left\{\begin{array}{ll}
e^{-\beta\mu\sqrt{-\tilde{g}_{tt}}},&(M-B)\\
\zeta_{\frac{D+1}{2}}[e^{-\beta\mu\sqrt{-\tilde{g}_{tt}}}],&(B-E)\\
-\zeta_{\frac{D+1}{2}}[-e^{-\beta\mu\sqrt{-\tilde{g}_{tt}}}],&(F-D)
\end{array}\right. \\ 
&\rightarrow&0, \hskip 0.5cm {\rm when} \hskip 0.5cm \beta\rightarrow\infty.
\end{eqnarray}

In the low temperature approximation, the thermodynamical potential and the
entropy $S=\beta^2\frac{\partial W}{\partial\beta}$ all exponentially tend 
to become zero suppose that $\tilde{g}_{tt}$ remains finite at every point 
in the space-time. They will be divergent at the point where $\tilde{g}_{tt}$
vanishes. This is physically reasonable and is consistent with the third 
law of the usual thermodynamics.

\vskip 8pt
\noindent
{\Large 6 Discussion: horizon and divergence}

We have stressed that whether the thermodynamical potential $W$ (or the free 
energy $F$) diverges or not apparently depends upon whether the metric tensor 
$\tilde{g}_{tt}$ (or ${g}_{tt}$) vanishes or not. The divergence appears if
and only if $\tilde{g}_{tt}$ is equal to zero. Although there exists a event
horizon on a black hole background, however if $\tilde{g}_{tt}$ is nonzero at
this horizon, then no divergence will appear. Thus a necessary and sufficient 
condition that the divergence appears at the horizon is that the dragged metric 
tensor $\tilde{g}_{tt}$ is equal to zero at this horizon. Because $\tilde{g}_{
tt}$ degenerates to ${g}_{tt}$ when the angular velocity $\Omega_0$ is zero, 
so we only need to study the general case, namely $\Omega_0\not=0$. 
 
Suppose $r_c$ is the $\rho$-fold root of equation: $\tilde{g}_{tt}=0$, then we 
can recast $\tilde{g}_{tt}$ around the point $r_c$ into the form
\begin{equation}
\tilde{g}_{tt}=(r-r_c)^{\rho}G(x)\approx (r-r_c)^{\rho}G(r_c), \hskip 0.5cm 
G(r_c)=\frac{1}{\rho !}\frac{d^{\rho}}{dr^{\rho}}\tilde{g}_{tt}(r_c)
=\frac{1}{\rho !}\tilde{g}_{tt}^{(\rho)}(r_c),
\end{equation}

\noindent
here $G(x)$ being an analytical function at the point $r_c$, the $(D-2)$-th 
coordinate $x^{D-2}=r$ is a "radial" coordinate. In the first one of the 
above-head equation, the second expression is obtained by taking the lowest 
order approximation. 

Actually the point $r_c$ is located at the light velocity surface$^{5,7-10}$. 
In the case $\Omega_0=0$, the surface such that $g_{tt}=0$ is the infinite
red-shift surface. Apparently, the location of the light velocity surface 
depends upon the choice of the angular velocity $\Omega_0$. Further, let us 
assume\footnote{In fact, this assumption is reasonable physically in the case 
of a black hole.} that the horizon is on the light velocity surface, namely, 
the location of the horizon satisfies equation $\tilde{g}_{tt}(r_h)=0$. Near 
the horizon $r_h$, the dragged metric tensor $\tilde{g}_{tt}$ tends to become 
zero.

From the asymptotic expression of $D/2$-th order MacDonald function $K_{D/2}(z)$
at small $z$, one can know that the thermodynamical potential near the horizon 
has the same behavior as it does in the high temperature approximation or in 
the zero mass case:
\begin{equation}
-W\approx\frac{\Gamma(D/2)}{\pi^{D/2}\beta^D}\int d^{D-2}xd\varphi
\sqrt{\frac{-g_D}{(-\tilde{g}_{tt})^D}}{\cal{N}}_D
={\cal{N}}_D\frac{\Gamma(D/2)\tilde{V}_{D-1}}{\pi^{D/2}\beta^D}.
\end{equation}

Substituting the lowest approximation of the metric tensor $\tilde{g}_{tt}$
into the expression of the volume of the dragged optical space, the leading 
behavior of $\tilde{V}_{D-1}$ near the horizon is given by
\begin{eqnarray}\nonumber
\tilde{V}_{D-1}&=&\int d^{D-2}xd\varphi\sqrt{\frac{-g_D}{(-\tilde{g}_{tt})^D}} 
\\  \nonumber
&\approx&\int d^{D-3}xd\varphi\int_{r_h+\epsilon}^Ldr(r-r_h)^{-D\rho/2}
\sqrt{\frac{-g_D}{(-G)^D}}(r_h) \\
&\approx&\frac{2}{D\rho-2}\epsilon^{1-D\rho/2}
\int d^{D-3}xd\varphi\sqrt{\frac{-g_D}{(-G)^D}}(r_h).
\end{eqnarray}

Here we introduce a small cut-off $\epsilon$ and another cut-off $L>>r_h$ to
remove the infra-red divergence and the U-V divergence, respectively. The 
leading behavior of the entropy near the horizon is given by
\begin{equation}
S\approx{\cal{N}}_D\frac{2D\Gamma(D/2)}{(D\rho-2)\pi^{D/2}\beta^{D-1}}
\epsilon^{1-D\rho/2}{\cal{F}}(r_h),
\end{equation}

\noindent
here ${\cal{F}}(r_h)=\int d^{D-3}xd\varphi\sqrt{\frac{-g_D}{(-G)^D}}(r_h)$ is 
proportional to the "area" of the event horizon.

The leading entropy of an idea gas obeying three kinds of the usual statistics
diverges in $\epsilon^{1-D\rho/2}$ as the system approaches the horizon of 
a black hole if and only if the location of the horizon is located at the
light velocity surface. Under such a circumstance, the leading behavior of
the entropy at temperature $1/\beta=\kappa/(2\pi)$ is proportional to the 
horizon "area", but diverges as the brick wall cut-off $\epsilon$ goes to 
zero. The divergence depends on the dimensional number $D$ of the space-time 
as well as the degeneracy $\rho$ of the horizon. The fundamental reason of the
divergence is that the volume of the dragged optical space tends to become 
infinite near the horizon which results in that the density of states for a 
given energy $E$ diverges as the system approaches the horizon.

However, although there doesn't exist a horizon in a non-black-hole space-time, 
the divergence will also appear when the system approaches the light velocity 
surface. Thus our conclusion is that the divergence has no direct relation
to the horizon and it is only determined by the equation of the light velocity
surface. A necessary condition that the divergence appears is that the
horizon is located at the light velocity surface. In the following section,
we will use some concrete examples to illustrate this viewpoint.

\vskip 8pt
\noindent
{\Large 7 Examples}

In this section, on the one hand, we will give some concrete examples to 
discuss the divergence problem, on the other hand, we will determine the
location of the event horizon and its surface gravity in a given black hole
geometry. 

\vskip 5pt
\noindent
{\large Example A: Minkowski space-time ($\alpha=0$)}

In the four dimensional flat geometry, $g_{tt}=-1, A_t=0, A_{\varphi}=0$, 
one can select a gauge that satisfies ${\cal{B}}=\Phi_0=0$. If a quantum field 
has a vanishing angular velocity $\Omega_0=0$, then the total number of 
states and the free energy are given by
\begin{eqnarray}
\Gamma({\cal{E}})&=&\frac{1}{(4\pi)^{3/2}\Gamma(5/2)}\int d^3x\sqrt{-g_4}
[{\cal{E}}^2-\mu^2]^{3/2}=\frac{V_3}{6\pi^2}[{\cal{E}}^2-\mu^2]^{3/2}, \\
-F&=&2V_3(\frac{\mu}{2\pi\beta})^2
\left\{\begin{array}{ll}
K_2(\mu\beta),&(M-B),\\
\sum\limits_{n=1}^{\infty}\frac{K_2(n\mu\beta)}{n^2},&(B-E),\\
\sum\limits_{n=1}^{\infty}\frac{(-1)^{n+1}K_2(n\mu\beta)}{n^2},&(F-D).
\end{array}\right. 
\end{eqnarray}

\noindent
Here the volume of the optical space $V_3=\int_{\rm box}d^3x\sqrt{-g_4}$ is
is equal to the volume of the box that confines the idea gas being considered.

In the low temperature approximation, the free energy and the entropy all 
tend to become zero in $e^{-\beta\mu}$. In the high temperature approximation, 
the asymptotic behavior of the entropy is given by
\begin{equation}
S(\Omega_0=0)\approx{\cal{N}}_4\frac{4V_3}{\pi^2\beta^3},
\end{equation}

\noindent
here for convenience, the statistical factor is introduced ${\cal{N}}_4=1,
\zeta_4(1), -\zeta_4(-1)$, for M-B, B-E, F-D statistics, respectively. The 
Riemann zeta constants are all known $\zeta_4(1)=\pi^4/90, \zeta_4(-1)=7/8
\zeta_4(1)=7\pi^4/720$.

In the spherical coordinates frame, the Minkowski metric is given by $ds^2=
-dt^2+dr^2+r^2(d\theta^2+\sin^2\theta d\varphi^2)$, the metric determinant is 
$g_4=-r^4\sin^2\theta$, the volume of optical space is equal to that of the 
box, namely, $V_3=4\pi\int\limits_0^Rr^2dr=\frac{4\pi}{3}R^3$.

No horizon exists in the flat space-time for there is no solution satisfying
${\cal{D}}=-r^2\sin^2\theta=0, g_4=-r^4\sin^2\theta\not=0$, so the indices 
$\alpha=0, \rho=0$, the latter due to $g_{tt}=-1\not=0$. To make states density 
and entropy finite, one must restrict the size of 3-dimensional sphere, and use 
a box to confine the propagation of a quantum field. The box acts as imposing 
a boundary condition on the quantum field. This will result in quantization 
of energy and discretization of phase space volume as well as state density.

However, when a quantum field has a non-zero angular velocity $\Omega_0\not=0$,
the dragged metric tensor is $\tilde{g}_{tt}=-1+r^2\sin^2\theta\Omega_0^2$. The 
thermodynamical potential is given by
\begin{eqnarray}
-W=2(\frac{\mu}{2\pi\beta^2})^2\int drd\theta d\varphi
\frac{r^2\sin\theta}{-\tilde{g}_{tt}}
\left\{\begin{array}{ll}
K_2(\mu\beta\sqrt{-\tilde{g}_{tt}}),&(M-B),\\
\sum\limits_{n=1}^{\infty}\frac{K_2(n\mu\beta\sqrt{-
\tilde{g}_{tt}})}{n^2},&(B-E),\\
\sum\limits_{n=1}^{\infty}\frac{(-1)^{n+1}K_2(n\mu\beta\sqrt{-
\tilde{g}_{tt}})}{n^2},&(F-D).
\end{array}\right. 
\end{eqnarray}

The entropy at small $\beta\mu\sqrt{-\tilde{g}_{tt}}$ behaves like
\begin{equation}
S(\Omega_0\not=0)\approx{\cal{N}}_4\frac{4\tilde{V}_3}{\pi^2\beta^3},
\end{equation}

\noindent
here the volume of the dragged optical space $\tilde{V}_3=2\pi\int drd\theta 
r^2\sin\theta(1-r^2\sin^2\theta\Omega_0^2)^{-2}$.

The entropy diverges as the angular velocity $|\Omega_0|\rightarrow (r\sin
\theta)^{-1}$. To preserve the volume of the dragged optical space and the
entropy finite and real, one must restrict the velocity $|\Omega_0|<(r\sin
\theta)^{-1}$. Otherwise the entropy will diverge or become imaginary. For a
finite $\Omega$, the light velocity surface is located at the surface that 
$r\sin\theta=\pm\Omega_0^{-1}$, then the index $\rho=1$. Under the condition 
$|\Omega_0r\sin\theta|<1$, the volume of the dragged optical space is
\begin{eqnarray}\nonumber
\tilde{V}_3&=&2\pi\sum\limits_{k=0}^{\infty}(k+1)\Omega_0^{2k}\int_0^Rr^{2k+2}dr
\int_0^{\pi}\sin^{2k+1}\theta d\theta \\ 
&=&4\pi R^3\{1+\sum\limits_{k=1}^{\infty}\frac{[(k+1)]!]^2}{(2k+2)!(k+3/2)}
(2\Omega_0R)^{2k}\}.
\end{eqnarray}

The condition that the power series in the above-head expression converges is
$|\Omega_0R|<1$. When the angular velocity $\Omega_0=0$, the dragged volume is
equal to that of the sphere $\tilde{V}_3^0=4\pi R^3/3$. When $\Omega_0\not=0$,
its zero-th order approximation is also equal to the sphere volume.

\vskip 5pt
\noindent
{\large Example B: Four dimensional static black hole ($\alpha=1;2$)}  

Next let us consider the Reissner-Nordstrom black hole, the metric and 
electro-magnetical potential are 
\begin{eqnarray}
ds^2&=&-\frac{\Delta}{r^2}dt^2+\frac{r^2}{\Delta}dr^2
+r^2(d\theta^2+\sin^2\theta d\varphi^2),\\ 
A&=&-\frac{Q}{r}dt, \hskip 0.8cm \Delta=r^2-2Mr+Q^2
\end{eqnarray}

The metric determinant is $g_4=-r^4\sin^2\theta$. Conditions ${\cal{D}}=-\Delta
\sin^2\theta=0, g_4=r^4\sin^2\theta\not=0$ can be satisfied by the horizon
surface equation $\Delta=0$. In the non-extremal case ($M^2\not=Q^2$), the
metric has two horizons at $r_{\pm}=M\pm(M^2-Q^2)^{1/2}$. Let $r_h=r_{\pm}$, 
then the horizon function can be rewritten as $\Delta=(r-r_h)[r-r_h+2(r_h-M)]
\approx 2(r-r_h)(r_h-M)$. So we have the index $\alpha=1$ and function $H(r_h)
=-2(r_h-M)\sin^2\theta$. 

To preserve the state density real and finite, we should restrict the system 
in the region that $\Delta>0$ due to $-g_{tt}>0$. Suppose that the angular 
velocity is zero, $\Omega_0=0$, and choose such a potential $\Phi_0=-A_t=Q/r$ 
that makes the chemical potential ${\cal{B}}=0$, then the light velocity surface 
coincides the the horizon due to $g_{tt}=-\Delta/r^2=0$. So we have the index
$\rho=1$ and constant $G(r_h)=-2(r_h-M)/r_h^2=-2\kappa_h$. The leading term of 
the volume of the optical space is given by 
\begin{eqnarray}
V_3=4\pi\int dr\frac{r^6}{\Delta^2}
\approx\int_{r_h+\epsilon}^Ldr\frac{\pi r_h^6}{(r-r_h)^2(r_h-M)^2} 
\approx\frac{\pi r_h^6}{\epsilon(r_h-M)^2}
=\frac{{\cal{A}}_h}{4\epsilon\kappa_h^2}, 
\end{eqnarray}
 
\noindent
here ${\cal{A}}_h=4\pi r_h^2$ is the area of the horizon with the surface
gravity being $\kappa_h=(r_h-M)/r_h^2$. In terms of the proper distance
cut-off from the horizon $r_h$ to $r_h+\epsilon$:
$$\delta=\int_{r_h}^{r_h+\epsilon}\Delta^{-1/2}rdr\approx\int_{r_h}^{r_h
+\epsilon}dr/\sqrt{2\kappa_h(r-r_h)}=\sqrt{2\epsilon/\kappa_h},$$

\noindent 
the volume of the optical space is rewritten as $V_3={\cal{A}}_h/(2\kappa_h^3
\delta^2)$. The leading behavior of the entropy near the horizon is:
\begin{equation}
S(\Omega_0=0)\approx{\cal{N}}_4\frac{2{\cal{A}}_h}{\pi^2(\beta\kappa_h)^3
\delta^2}.
\end{equation}

At the Hartle-Hawking temperature $\beta=\kappa_h/(2\pi)$, the entropy of an 
idea gas is proportional to the horizon area ${\cal{A}}_h$ and diverges in 
$\delta^{-2}$:
\begin{equation}
S(\Omega_0=0)\approx{\cal{N}}_4\frac{{\cal{A}}_h}{4\pi^5\delta^2}.
\end{equation}
	 
In the case of the extremal Reissner-Nordstrom black hole ($M^2=Q^2$), the
horizon location $r_h=M$ is the double root of equation $\Delta=(r-M)^2=0$,
so the indices $\alpha=\rho=2$. As the function ${\cal{D}}=-(r-M)^2\sin^2\theta, 
g_{tt}=-(r-M)^2/r^2\approx-(r-M)^2M^{-2}$, so we have constant $G(r_h)=-M^{-2}
$ and function $H(r_h)=-\sin^2\theta$. The leading term of the volume of the
optical space near the horizon is given by
\begin{eqnarray}
V_3=\int 4\pi dr\frac{r^6}{\Delta^2}
\approx\int_{M+\epsilon}^Ldr\frac{4\pi M^6}{(r-M)^4} 
\approx\frac{4\pi M^6}{3\epsilon^3}. 
\end{eqnarray}

Near the horizon, the entropy diverges cubically (in $\epsilon^{-3}$):
\begin{equation}
S(\Omega_0=0)\approx{\cal{N}}_4\frac{16M^6}{3\pi(\beta\epsilon)^3}.
\end{equation}

The entropy of a quantum field in the extremal Reissner-Nordstrom black hole
is proportional to the horizon area ${\cal{A}}_e=4\pi M^2$ only when the
temperature of the system $\beta\sim M^{4/3}$. The reason why the 
thermodynamics of the extremal black hole is ill-defined, however, is 
still unclear$^{15}$.   

\vskip 5pt
\noindent
{Example C: Four dimensional stationary axisymmetry black hole ($\alpha=1$)}

The third example in which we have an interest is the Kerr-Newman black hole.
In the Boyer-Lindquist coordinates, the metric and the electro-magnetical 
potential of the Kerr-Newman black hole takes the form    
\begin{eqnarray}\nonumber
ds^2&=&-\frac{\Delta-a^2\sin^2\theta}{\Sigma}dt^2-2\frac{r^2+a^2
-\Delta}{\Sigma}a\sin^2\theta dtd\varphi \\
& &+\frac{(r^2+a^2)^2-\Delta a^2\sin^2\theta}{\Sigma}\sin^2\theta 
d\varphi^2+\Sigma(\frac{dr^2}{\Delta}+d\theta^2), \\
A&=&-\frac{Qr}{\Sigma}(dt-a\sin^2\theta d\varphi)
\end{eqnarray}

\noindent
with the event horizon function $\Delta=r^2-2Mr+Q^2+a^2, \Sigma=r^2+a^2\cos^2
\theta, {\cal{D}}=-\Delta\sin^2\theta$, and the metric determinant $ g_4=-
(\Sigma\sin\theta)^2$. The location of the horizon $r_h=r_{\pm}=M\pm(M^2-Q^2
-a^2)^{1/2}$ satisfies conditions ${\cal{D}}=0, g_4\not=0$ by equation $\Delta
=0$. For the non-extremal case, the index is equal to one ($\alpha=1$), and
function $H(r_h)=-2(r_h-M)\sin^2\theta$.

The entropy of a quantum scalar field in the non-extreme Kerr-Newman black 
hole had been considered by many authors$^{7-9,16}$ in the case that the
scalar field is co-rotating with the black hole, namely the angular velocity 
is a constant $\Omega_h=a/(r_h^2+a^2)$, and the potential $\Phi_h=Qr_h/(r_h^2
+a^2)$, thus the chemical potential tends to become zero near the horizon.
Other than this choice and the trial choice $\Omega_0=0$, however, we choose 
a local angular velocity and a local potential as:
\begin{eqnarray}\nonumber
\Omega_0&=&\Omega=-\frac{g_{t\varphi}}{g_{\varphi\varphi}}=\frac{a(r^2+a^2
-\Delta)}{(r^2+a^2)^2-\Delta a^2\sin^2\theta},\\
\Phi_0&=&-(A_t+\Omega A_{\varphi})=\frac{Qr(r^2+a^2)}{(r^2+a^2)^2
-\Delta a^2\sin^2\theta}.
\end{eqnarray}

The reason why we choose a local velocity and a local potential is that the 
light velocity surface depends upon the choice of the angular velocity. For
instance, in the case of $\Omega_0=0$, the points satisfying $\tilde{g}_{tt}
=0$ are on the stationary limit surface. The local velocity and potential on 
the horizon are $\Omega_0=\Omega_h, \Phi_0=\Phi_h$, respectively. They tends 
to become zero at infinity. Under such a choice, the chemical potential is 
always equal to zero ($B=0$), and the dragged metric tensor becomes:
$$\tilde{g}_{tt}=g_{tt}+2g_{t\varphi}\Omega+g_{\varphi\varphi}\Omega^2
=1/g^{tt}=-\Delta\Sigma/[(r^2+a^2)^2-\Delta a^2\sin^2\theta].$$

The light velocity surface coincides with the horizon under our choice because 
the equation $\tilde{g}_{tt}=1/g^{tt}=0$ can be satisfied by $\Delta=0$. Thus
the index $\rho=1$. To prevent from the presence of infinite and imaginary
state density, one must restrict the system satisfying inequalities: $(r^2+a^2
)^2>\Delta a^2\sin^2\theta>0$. Also, it places a lower bound and an upper bound 
on the angular velocity $\Omega_0$: $\Omega-\sqrt{-{\cal{D}}g_{\varphi
\varphi}^{-2}}<\Omega_0<\Omega+\sqrt{-{\cal{D}}g_{\varphi\varphi}^{-2}}$. If 
only $\Delta>0$, then the choice of the velocity $\Omega_0=\Omega$ certainly 
satisfies this restrictions.

The leading term of the volume of the dragged optical space is given by
\begin{eqnarray}\nonumber
\tilde{V}_3&=&\int_0^{\pi}d\theta d\varphi\int_{r_h+\epsilon}^Ldr
\sqrt{-g_4}(\frac{{g}_{\varphi\varphi}}{-{\cal{D}}})^2 
=2\pi\int_0^{\pi}d\theta\int_{r_h+\epsilon}^Ldr\frac{\sin\theta}{\Sigma
\Delta^2}[(r^2+a^2)^2-\Delta a^2\sin^2\theta]^2  \\  \nonumber
&\approx&\int_0^{\pi}\frac{\sin\theta d\theta}{r_h^2+a^2\cos^2\theta}
\int_{r_h+\epsilon}^Ldr\frac{\pi(r_h^2+a^2)^4}{2(r_h-M)^2(r-r_h)^2} 
\approx\frac{{\cal{A}}_h}{4\epsilon\kappa_h^2}\times
\frac{r_h^2+a^2}{ar_h}\arctan\frac{a}{r_h}.
\end{eqnarray}

The proper distance from $r_h$ to $r_h+\epsilon$ is a function of $\theta$:
$$\delta=\int_{r_h}^{r_h+\epsilon}\sqrt{\frac{\Sigma}{\Delta}}dr
\approx\int_{r_h}^{r_h+\epsilon}dr\sqrt{\frac{r_h^2+a^2\cos^2\theta}{2
(r_h-M)(r-r_h)}}
\approx\sqrt{\frac{2\epsilon(r_h^2+a^2\cos^2\theta)}{r_h-M}}.$$

In terms of the proper distance cut-off $\delta$, the dragged optical volume 
can be rewritten as
\begin{eqnarray}\nonumber
\tilde{V}_3\approx\frac{{\cal{A}}_h}{2\kappa_h^3\delta^2}\times
\frac{r_h^2+a^2\cos^2\theta}{ar_h}\arctan\frac{a}{r_h}
\approx\frac{{\cal{A}}_h}{2\kappa_h^3\delta^2}, \hskip 0.5cm (a<<r_h).
\end{eqnarray}
 
\noindent
Here ${\cal{A}}_h=4\pi(r_h^2+a^2)$ is the area of the event horizon, and 
$\kappa_h=(r_h-M)/((r_h^2+a^2)$ is the surface gravity. In the second 
approximation, we have taken the slow rotating limit. The leading behavior 
of the entropy near the horizon is:
\begin{equation}
S\approx{\cal{N}}_4\frac{2{\cal{A}}_h}{\pi^2(\beta\kappa_h)^3
\delta^2}.
\end{equation}

If we take $1/\beta$ as the Hartle-Hawking temperature $\kappa_h/(2\pi)$, the 
entropy of an idea gas near the horizon is proportional to the horizon area 
${\cal{A}}_h$ and diverges in $\delta^{-2}$ as $\delta\rightarrow 0$:
\begin{equation}
S\approx{\cal{N}}_4\frac{{\cal{A}}_h}{4\pi^5\delta^2}.
\end{equation}
	 
This leading behavior of the entropy of quantum fields near the horizon is 
a general form in the black hole background. It is proportional to the horizon 
area but diverges as the system approaches to the horizon. The reason of the 
divergence is due to the infinite state density for a given energy near the 
horizon. This agrees with the conclusions in Ref. 8 and 9.

As a brief summary, we has used the first example to demonstrate that 
although there exists no horizon in the flat space-time, if a quantum field 
has a vanishing angular velocity, one must also impose a box on the system to 
prevent from the presence of infinite state density. In the case that a 
quantum field has a non-vanishing angular velocity, the entropy also diverges 
as the system approaches to the light velocity surface. To guarantee the state 
density finite and real, the angular velocity must be restricted in a certain 
region.

The second example has been used to show that the thermodynamics of the 
extremal black hole is very different from that of the non-extremal black 
hole. In the third one, we have chosen a local angular velocity and a local
potential other than the popular uniform velocity. Both examples have shown
that the leading behavior of the entropy of a relativistical idea gas near 
the horizon is proportional to the horizon area and diverges as the system
approaches the horizon provided that the horizon is on the light velocity
surface.  
 
All examples have illustrated that the entropy of a quantum field is
proportional to the volume of the optical space or that of the dragged
optical space. In the case of a black hole, it is proportional to the
horizon area only after introducing a brick wall cut-off. In the four
dimensional black holes, the leading term of the entropy has a common
character.

Other cases can also be considered. In a lower than four dimensional
space-time, we have assumed that the statistics are the usual ones. However,
this may be problematical. In 2+1 dimensional planar system, anyons obeys a 
novel fractional statistics$^{20}$, neither the common Bose-Einstein statistics
nor the well-known Fermi-Dirac statistics. Theoretically, fractional quantum
Hall effect probably be interpreted by anyonic statistics$^{21}$. 

\vskip 8pt
\noindent
{\Large 8 Conclusion}

To summarize, a general framework of general relativistical thermodynamics 
for three kinds of the usual statistics has been done in a $D$-dimensional 
stationary axisymmetry space-time. We start from calculating the density of 
single particle by the classical phase space method. The density of single
state is invariant under a gauge transformation, however, it is suffered by 
the dragging of the angular velocity. To proceed, we assume that it is 
effective in an arbitrary dimensional space-time for a relativistical idea 
gas obeying the usual Maxwell-Boltzmann, Bose-Einstein or Fermi-Dirac 
statistics. A particular needed notice is that this assumption is probably 
invalid in a space-time with its dimensional number lower than four. In a 
space-time higher than the usual four dimension, no such problem exists.

Thermodynamical quantities such as the free energy or the thermodynamical 
potential and the entropy of a quantum field are evaluated. Exact analytical 
expressions for the free energy or the thermodynamical potential are in terms 
of the modified Bessel functions. In the high temperature approximation, the 
statistical entropy of a quantum field is proportional to the volume of optical
space in the case of a vanishing angular velocity of the quantum field. This 
conclusion agrees with the results that the entropy of a quantum field in a 
static space-time background is obtained by heat kernel expansion in the same 
approximation in Ref. 15. In the case that a quantum has a non-vanishing 
angular velocity, the entropy is proportional to the volume of the dragged 
optical space. In the low temperature approximation, the entropy tends to 
become zero exponentially.    

In general, the entropy of a quantum field in a $D$-dimensional space-time
depends upon the temporal component of the metric tensor $g_{tt}$ or the 
dragged metric tensor $\tilde{g}_{tt}$ only, as well as the dimensional number 
$D$. If the dragged metric tensor $\tilde{g}_{tt}$ doesn't vanish at every
point of the space-time being considered, then the entropy is finite provided  
the system is confined by a box. In the case of a black hole, the leading term 
of the entropy near the horizon is proportional to the horizon "area"  only 
when the horizon is located at the light velocity surface. The presence of 
horizon has no direct relation to the divergence of entropy, but it introduces 
an additional brick wall cut-off in place of the restrictions of a box. 
Although there exists a horizon of the black hole, the behavior of the entropy
near the horizon will also be finite if the light velocity surface doesn't 
coincides with the horizon. The divergence near the horizon has a definite 
relation to the dimensional number of a space-time.

As examples, we have discussed the four dimensional entropy of flat space-time
and that of black holes. The results agrees with the already-known results in 
the literatures. Using our general formation, one can compare the behaviors 
among the entropies in different dimensional space-times. It might not be a toy 
of everything, however, we wish it would work at least in a higher dimensional
space-time. In the case of a bosonic field, we don't consider such things as 
the contribution to the entropy from the superradiant modes and a probable 
existing phenomena of the well-known Bose-Einstein condensation here. In the 
process of our calculation of the free energy or the thermodynamical potential,
we have only made a small fugacity expansion also. however, we expect to 
discuss them in other places.

\vskip 8pt
\noindent
{\bf Acknowledgment}
This work is supported partially by the NSFC and Hubei-NSF in China.

\vskip 0.5cm
\begin{enumerate}
 \item J.D. Benkenstein, {\sl Do we understand black hole entropy?}, 
gr-qc/9409015. 
 \item J. Makela, P. Repo, gr-qc/9812075.
 \item S. Liberati, Nuovo Cimento, 112{\bf B}, 405 (1997), gr-qc/9601032.
 \item G. 't Hooft, Nucl. Phys. {\bf B}256, 727 (1985).
 \item H.C. Kim, M.H. Lee, and J.Y. Ji, gr-qc/9709062.
 \item A. Ghosh and P. Mitra, Phys. Rev. Lett. {\bf 73}, 2521 (1994).
 \item M.H. Lee and J.K. Kim, Phys. Lett. {\bf A}212, 323 (1996), hep-th/960212.
 \item M.H. Lee and J.K. Kim, Phys. Rev. {\bf D}54, 3904 (1996), hep-th/9603055.
 \item M.H. Lee and J.K. Kim, hep-th/9604130.
 \item M.H. Lee, H.C. Kim, and J.K. Kim, Phys. Lett. {\bf B}388, 487 (1996),
hep-th/9603072.
 \item J. Ho, W.T. Kim, Y.J. Park, and H. Shin, Phys. Lett. {\bf B}392, 311 
(1997), hep-th/9603043.
 \item S. Chandrasekhar, {\sl The Mathematical Theory of Black Hole}, 
(Clarendon, Oxford, 1983).
 \item J. Ho, W.T. Kim, Y.J. Park, and H. Shin, Class. Quant. Grav. {\bf 14},
2617 (1997), gr-qc/9704032. 
 \item J. Ho and G. Kang, Phys. Lett. {\bf B}445, 27 (1998), gr-qc/9806118.
 \item S.P. de Alwis and N. Ohta, hep-th/9412027; Phys. Rev. {\bf D}52, 3529
(1995), hep-th/9504033.
 \item J.L. Jing and M.L. Yan, gr-qc/9904001, to appear in Phys. Rev.{\bf
D}; gr-qc/9907011.
 \item L.E. Reichl, {\sl A Modern Course in Statistical Physics}, (Texas 
University Press, 1980).
 \item V.P. Frolov and D.V. Fursaev, gr-qc/9907046.
 \item Z. Zhao and Y.X. Gui, Acta Astrophysica Sinica, {\bf 3}, 146 (1983).
 \item S. Forte, Rev. Mod. Phys. {\bf 64}, 193 (1992); Y.S. Wu, Phys. Rev.
Lett. {\bf 52}, 2103 (1984); {\bf 53}, 111 (1984).
 \item F. Wilczek, {\sl Fractional Statistics and Anyon Superconductivity},
(World Scientific Press, 1990).
\end{enumerate}

\end{document}